\documentclass[11pt,twoside]{article}
\usepackage{asp2010}
\usepackage{epsf}
\usepackage{psfig}
\usepackage{graphicx}
\resetcounters

\begin{document}

\title{The Upper Initial Mass Function from Ultraviolet Spectral Lines}
\author{Claus Leitherer
\affil{Space Telescope Science Institute, 3700 San Martin Dr., Baltimore, MD 21218, USA}}

\begin{abstract}
The space-ultraviolet wavelength region contains strong spectral lines from massive, hot stars. These features form in winds and are sensitive to luminosity and mass, and ultimately provide constraints on the initial mass function. New radiation-hydrodynamical models of stellar winds are used to construct a theoretical spectral library of massive stars for inclusion in population synthesis. The models are compared to observations of nearby star clusters, of starburst regions in local galaxies, and of distant star-forming galaxies. The data are consistent with a near-universal Salpeter-type initial mass function. We find no evidence of environmental effects on the initial mass function. Some model deficiencies are identified: stellar rotation and binary evolution are not accounted for and may become increasingly important in metal-poor systems.         
\end{abstract}

\section{Detecting Elusive Massive Stars}

Despite their luminosity, massive stars pose severe challenges to their direct observation in the integrated spectra of extragalactic systems. Both thermodynamics and atomic physics conspire against their detection: owing to their high temperatures, massive stars emit most of their radiation in the extreme ultraviolet (UV), and they have few strong spectral lines in the optical wavelength region. As a result, we have little observational leverage on the O-star population in the visual. The common remedy is to use nebular H{$\alpha$ emission as a proxy for the stellar ionizing radiation. In principle, this technique is a direct measure of the number of massive stars (Kennicutt 1998) but it is prone to often large uncertainties because the stellar UV photons have path lengths of tens of pc before being absorbed by interstellar gas. A superior method of counting the hard UV photons makes use of stellar winds: massive stars lose mass in powerful winds driven by radiation pressure. The process is well understood observationally and theoretically and serves as the basis of a technique to observe the strong P Cygni-type wind features in the satellite-UV. These features depend on the wind properties, which in turn depend on the stellar luminosity $L$. Since there is a well established mass-luminosity relation, the wind lines trace stellar mass and the initial mass function (IMF) of the population (Leitherer 2010). This is equivalent to the nebular H$\alpha$ method but circumvents the interstellar radiative transfer in gas and dust because the wind absorption occurs in the outskirts of the star itself. {\em In fact, this is the only viable method to conduct a census of massive stars in integrated spectra by their direct detection.}

\section{New Ultraviolet Models for Young, Massive Populations}

Our goal is the interpretation of the observed spectral features of massive stars in the satellite-UV. We employ the well-established population synthesis technique (Brinchmann 2010) for this purpose. This technique relies on several parameters, models, and assumptions: (i) A spectral library is needed. This library can be empirical or theoretical. Here I will make a pitch for a new theoretical library we have developed for the specific purpose of modeling the UV features of massive stars (Leitherer et al. 2010). (ii) Stellar evolution models provide the mass-luminosity relation and set the time scale. In the case of low-mass stars, stellar evolution is rather well understood. The situation is very different for massive stars where major uncertainties still prevail. Very often, stellar evolution models are the weak link in population synthesis models of massive stars. Most of the results I am reporting here rely on stars still relatively close to the main-sequence where the uncertainties are smaller. (iii) Parameters such as dust reddening, chemical composition, etc., are not treated self-consistently here but are considered as input from observations. (iv) The only free parameters in the modeling are the star-formation history of the population and the IMF. Since massive stars have evolutionary timescales of less than $\sim$50~Myr, star-formation equilibrium is reached quickly, and the population properties become time-independent in statistically large systems. {\em This essentially leaves the IMF as the only free parameter.} 

Prior attempts of modeling the UV spectra of galaxies using libraries were quite successful (e.g., Pettini et al. 2000) but the limitations are obvious: the lack of spectra
of metal-poor massive stars, the presence of $\alpha$-element/Fe variations, and the generally low S/N of the UV spectra of the template stars. A first effort to model UV spectra purely theoretically was done by Rix et al. (2004), who matched the weak photospheric features in the spectra to determine stellar chemical abundances.
The state of the stellar atmospheres at that time precluded the modeling of the stellar-wind lines in the spectra, which was frustrating because the wind lines are the strongest
spectral features and can be detected even in low-S/N spectra of distant galaxies. Substantial astrophysical (e.g., inclusion of wind structure and X-ray ionization) and
technical (e.g., decrease of computing time by an order of magnitude) improvements in atmospheric modeling are now allowing us to compute the full spectrum, including the
wind lines (Leitherer et al. 2010). We used WM-Basic, a non-LTE, spherically extended, blanketed, radiation-hydrodynamics code for hot stars (Pauldrach et al. 2001) for a self-consistent
calculation of the photospheric and wind parameters and the generation of synthetic spectra between 900 and 3000~\AA. The library covers the relevant parameter space of hot stars 
with masses $M > 5$~M$_\odot$ on and beyond the main-sequence for abundances between 2$\times$ and 0.05$\times$~Z$_\odot$. The spectral resolution is 0.4~\AA. An example is shown in Fig.~1 where we reproduce two representative spectra. The hot star (left panel) has a spectral type of O6~III. Prominent spectral lines are identified. Most of these lines are formed in the stellar wind, as can be seen from the blueshifted absorption components with velocities exceeding $\sim$2000~km~s$^{-1}$. S~VI $\lambda$939, O~VI $\lambda$1035, P~V $\lambda$1123, N~V $\lambda$1240, and C~IV $\lambda$1550 are the strongest features that are uniformly present in O stars. Si~IV $\lambda$1400 is weak in this particular model but can become a strong line in supergiants whose denser winds lead to recombination from Si$^{4+}$ (which is the dominant ionization stage) to Si$^{3+}$ (Walborn \& Panek 1985; Drew 1989). Other features present in this spectrum are often not detectable in the spectrum of a typical stellar population whose numerous B stars do not show these lines and dilute the O star contribution. The wavelength region longward of 1800~\AA\ is devoid of both stellar-wind and photospheric lines. This region has little diagnostic value for constraining an O-star population. The O-star spectrum in Fig.~1 can be contrasted with the B-star spectrum in the right panel which corresponds to a B3~Ia supergiant. The high-ionization lines present in the O-star spectrum are much weaker or completely absent in B stars. The strongest lines are, among others, C~III $\lambda$1176, C~II $\lambda$1335, Si~IV $\lambda$1400, and Fe~III $\lambda$1893. This star has a terminal wind velocity of $v_\infty = 300$~km~s$^{-1}$, resulting in comparatively small blueshifts of the absorption components. The line-blanketing by photospheric absorption lines is significant, in particular at the shortest wavelengths where the rectified flux is close to the zero level.

\begin{figure}
\epsscale{10.5}
\plottwo{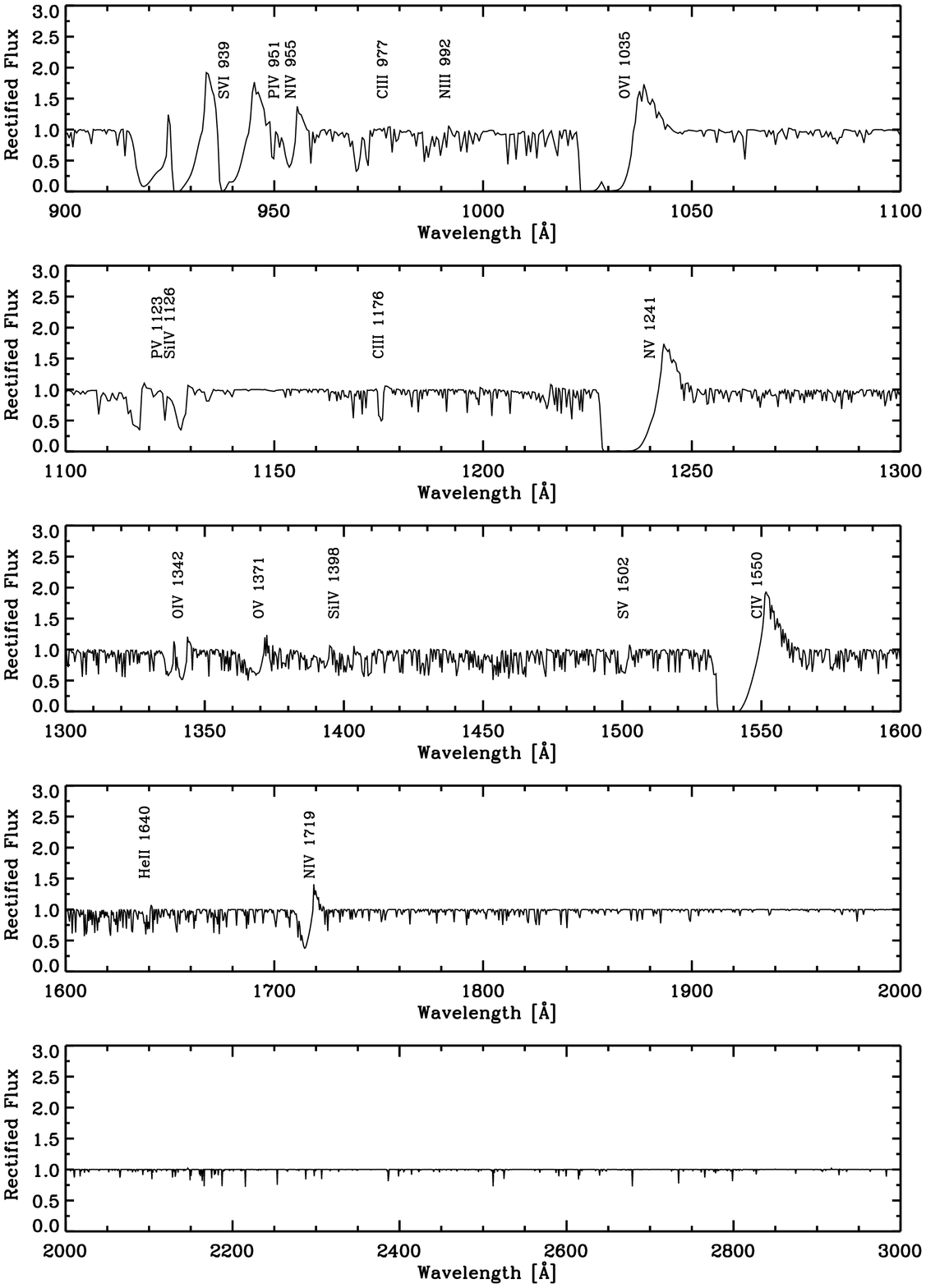}{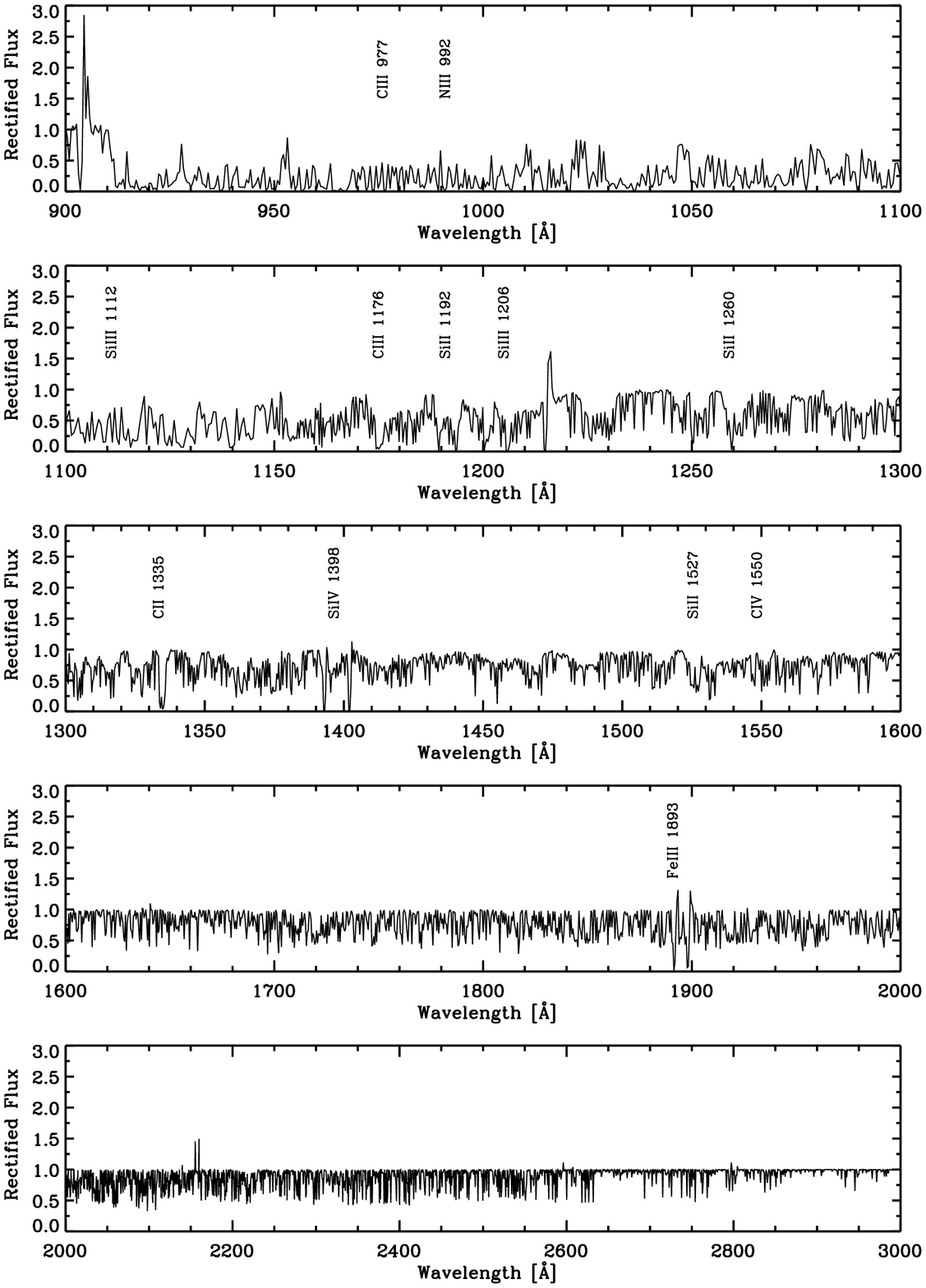}
\caption{Examples of individual spectra in the new UV library. Left panel: a hot O6~III star; right panel: a cool, luminous B3~Ia star. The strongest spectral features are labeled. From Leitherer et al. (2010).}
\end{figure}
We implemented the stellar library in Starburst99 (Leitherer et al. 1999; Leitherer \& Chen 2009) where it complements and extends the empirical libraries currently in use. We tested the new library both at the individual star and at the stellar population level and found no significant deficiencies in the new models. Therefore the new theoretical library is the default in Starburst99 and allows us to model and interpret the UV spectra of massive star populations for essentially any chemical composition of interest and at any redshift.

\section{Probing the Initial Mass Function}

Local star-forming galaxies targeted for UV spectroscopy are necessarily UV-bright, a bias imposed by the low quantum efficiency of UV detectors and the relatively small telescope sizes. Morphologically, these galaxies tend to be of late Hubble types. They include blue compact galaxies, H II galaxies, and nuclear starbursts. The stellar masses are of order $10^9$~M$_\odot$, and the absolute magnitudes are in the range $-16 > M_{\rm B} > -19$. Oxygen abundances are as low as 1/20 Z$_\odot$ and as high as Z$_\odot$, with typical values similar to those of the Magellanic Clouds. The first scientifically useful UV spectra of star-forming galaxies were obtained with the IUE satellite but these data were quickly superseded by higher resolution ($R > 1000$) and higher S/N ($>10$) spectra collected mostly with HST's first- (FOS, GHRS), second- (STIS), and third-generation (COS) spectrographs.

The different aperture sizes of the HST spectrographs (together with the range of galaxy distances) result in a broad range of spatial scales sampled by the observations. Narrow-slit data typically cover a few pc and encompass an individual massive star cluster. In this case, the UV lines are degenerate with respect to age and IMF, and the determination of one of the two requires independent knowledge of the  other. For instance, the detection of Wolf-Rayet (W-R) features or the properties of a surrounding H~II region can be used for an age estimate. Comprehensive studies with STIS were done by, e.g., Tremonti et al. (2001), Gonz\'alez Delgado et al. (2002), Chandar et al. (2003, 2004, 2005), and Wofford et al. (2010) who uniformly found IMF slopes similar to the classical Salpeter value ($\alpha = 2.35$). No dependence on the environment was detected. In particular, the chemical composition of the H~II region does not appear to influence the IMF. See also A. Wofford's contribution in these proceedings.

Recently we mined the HST data archive and extracted and processed all scientifically useful UV spectra of star-forming galaxies obtained with FOS and GHRS (Leitherer et al. 2011). This resulted in 46 UV spectra of 28 galaxies ranging in distance from 2 to 250~Mpc. The angular sizes of the FOS and GHRS apertures ($1'' - 2''$) are an order of magnitude smaller than that of IUE, yet they are an order of magnitude larger than those used with STIS (typically $0.2''$). As opposed to the STIS data, FOS and GHRS spectra encompass small starburst regions with sizes of $\sim$100~pc (corresponding, e.g., to $2''$ at 10~Mpc). This is large enough to expect a significant spread in stellar population ages. As a result, the UV spectra can be assumed to be age independent. In Fig.~2 we show the average of all 46 individual spectra. This high-S/N spectrum can be considered as representative for the average UV spectrum of star-forming galaxies in the local universe. The average oxygen abundance of log(O/H) + 12 = 8.5 is close to, but slightly below the solar value. We generated a synthetic UV spectrum for a standard equilibrium population at solar chemical composition following a Kroupa IMF between 0.1 and 100~M$_\odot$. (Note that this IMF is identical to the Salpeter IMF for massive stars.) The comparison with the data in Fig.~2 suggests excellent agreement. The key constraints are the N~V, Si~IV, and C~IV lines. The interpretation of the Si~IV doublet requires care because of the presence of additional strong interstellar lines which are obviously absent in the model. We can exclude IMF slopes steeper than $\sim$2.6 and flatter than $\sim$2.0. {\em There are no additional adjustable parameters}. We also divided the sample into sub-solar and super-solar abundance sub-samples to search for evidence of IMF variations. No significant departure from the standard Kroupa IMF was found.

\begin{figure}
\includegraphics[angle=90,width=5.25in]{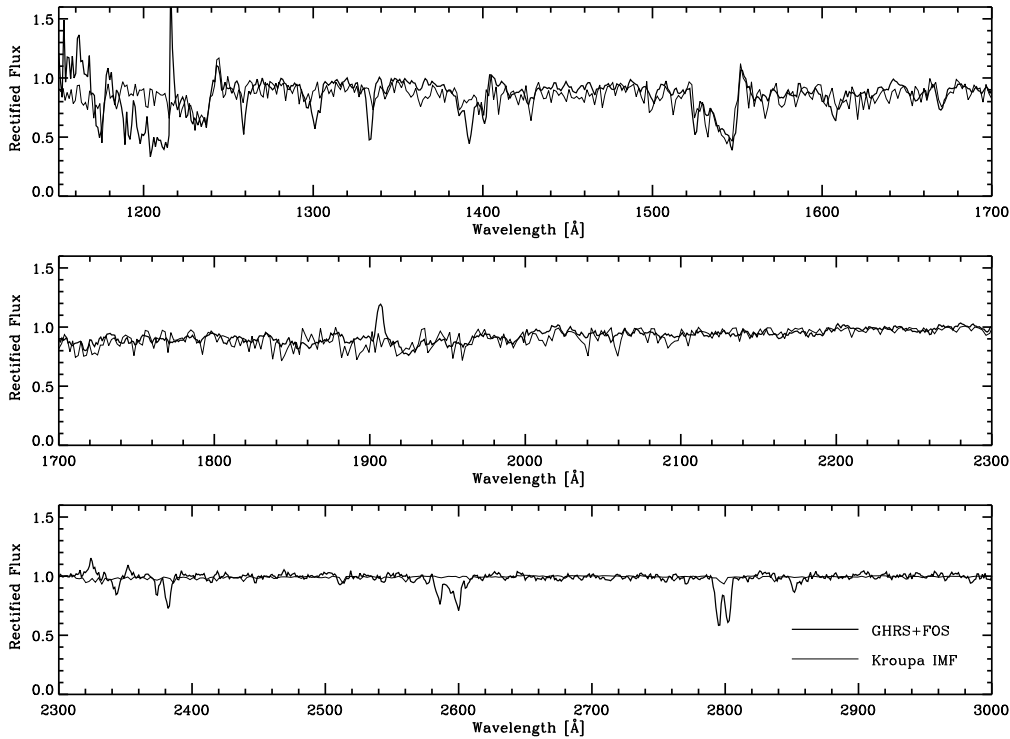}
\caption{Comparison between the average spectrum of 46 local star-forming regions generated by Leitherer et al. (2011; thick line) and a synthetic spectrum with a Kroupa/Salpeter IMF populated with stars up to $M = 100$~M$_\odot$ (thin line). The narrow absorption lines (e.g., at 1260~\AA, 1335~\AA, 2600~\AA, 2800~\AA) are interstellar and absent in the synthetic spectrum. There is strong geocoronal and Milky Way Ly-$\alpha$ contamination in the data. Note the agreement for the wind features of N~V $\lambda$1240 and C~IV $\lambda$1550.  
}
\end{figure}

The new generation of 8 to 10-m class ground-based telescopes can produce restframe-UV spectra of star-forming galaxies at cosmological distances (often referred to as ``Lyman-break galaxies''; LBGs) whose quality rivals and often exceeds those of their local counterparts (e.g., Pettini 2008; Quider et al. 2009, 2010). A comparison of the spectra of local and distant star-forming galaxies suggests striking similarity. While the two samples are similar in many respects, it is important to realize that the average luminosity and mass of the high-redshift sample are 1 to 2 orders of magnitude higher than for their local counterparts. What do the UV lines observed at high redshift tell us about the IMF in these galaxies? Shapley et al. (2003) performed a systematic study of the restframe UV spectroscopic properties of LBGs. Their database of $\sim$1000 LBG spectra proves useful for constructing a high S/N composite spectrum. The wavelength region around the C~IV $\lambda$1550 line is compared with models in Fig.~3. C~IV is observed as a P~Cygni profile but with a deep absorption whose origin is likely interstellar. Interstellar C~IV (and other interstellar lines) in LBGs is comparatively stronger than in local star-forming galaxies. This difference reflects different conditions in the velocity dispersion and/or the covering fraction of the interstellar gas. Comparing only the emission part and the blue wing of the {\em stellar} C~IV with the models, good agreement is found with the model obtained for an IMF slope of $\alpha = 2.5$. The steeper and the flatter IMF in Fig.~3 are not consistent with the observed C~IV line. $\alpha = 2.5$ is slightly steeper than the Salpeter value found in the local sample but the statistical significance of this difference is questionable. Taken at face value, it might reflect an aperture size effect: at high redshift, a large portion of the older population in the galaxy disk is sampled, whose IMF may differ from the IMF of young starburst clusters. We may observe the difference between the IMF and the IGIMF (Kroupa (2008).

\begin{figure}
\plotone{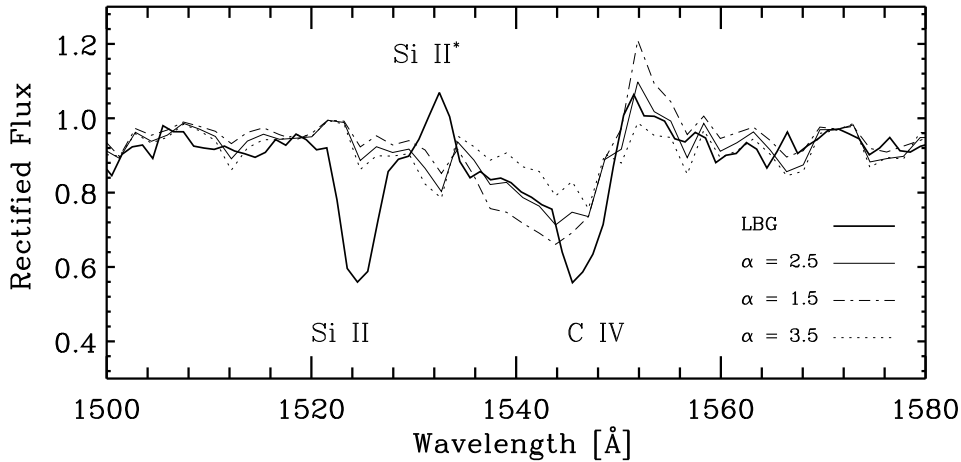}
\caption{Comparison of the composite LBG spectrum of Shapley et al. (2003; thick line) and three synthetic models with IMF slopes between 1.5 and 3.5. The strong interstellar lines of Si~II $\lambda$1526 and Si~II$^*$ $\lambda$1533 are marked. C~IV $\lambda$1550 is stellar with an additional interstellar component.
}
\end{figure}

\section{Outliers and Open Issues}

The strength of the stellar UV features provides constraints on the relative proportion of the most massive O- relative to less massive B-stars. Care must be taken when using this information to derive the IMF. A cautionary tale is the apparent dearth of massive stars in the intercluster field population suggested by UV spectra  (Chandar et al. 2005). Long-slit spectroscopy of star-forming galaxies with HST's STIS allows separate studies of both the cluster and the intercluster light, which is the diffuse, unresolved stellar emission. Comparison of the two suggests weaker stellar N~V, Si~IV, and C~IV lines in the field. This translates into a deficit of very massive field stars. One interpretation could be a steeper field IMF, which has fewer O stars. More likely, an age effect could be responsible: Field stars are older on average because they are the relics of dissolved star clusters. The cluster lifetimes are longer than O-star lifetimes. Therefore, massive O stars will end their lives as cluster members whereas less massive, longer lived B stars will diffuse into the field when star clusters dissolve. This will mimic a different cluster vs. field IMF in UV spectra whereas the root cause is the complex dynamical evolution of the star-forming complexes.

The diagnostics I discussed so far avoid relying on very evolved stellar phases, such as W-R stars, which are the hot, helium-rich descendants of very massive O stars (Crowther 2007). Observations of local starbursts sometimes show weak, broad He~II $\lambda$1640 emission, which is indicative of W-R stars. This line is normally weak in local starbursts, as expected for a normal IMF and given the short W-R lifetimes. Chandar et al. (2004) observed extraordinarily strong He~II $\lambda$1640 emission in a star cluster in the dwarf galaxy NGC~3125. The equivalent width of 7.4~\AA\ is by far the largest value observed in the local universe. Understanding the UV spectrum required extreme assumptions, including an IMF heavily biased towards massive stars. Complicating the interpretation, however, were uncertainties in the stellar evolution modeling as well as the treatment of dust reddening (Hadfield \& Crowther 2006). Interestingly, broad (non-nebular) He~II $\lambda$1640 is the strongest stellar line in the composite LBG spectrum of Shapley et al. (2003). It is the most discrepant stellar line when the restframe UV spectra of LBGs are compared to UV templates of local starbursts such as those in Fig. 2. If W-R clusters were present in large numbers in LBGs, the LBG restframe UV spectra could be empirically reproduced and understood. This leaves open the question why such unique clusters are rare in the local universe and why they could be abundant in the early universe. 

Erb et al. (2010) discuss an LBG with even more extreme He~II emission. Q2343-BX418 is an $L^*$ galaxy at a redshift of 2.3 with a very low mass-to-light ratio and unusual properties: BX418 is young ($<100$~Myr), has low mass ($M \approx 10^9$~M$_\odot$), and low metallicity (log(O/H + 12 = 7.9). The restframe UV spectrum contains strong high-ionization interstellar absorption lines from outflowing gas, while the low-ionization lines are extremely weak. Erb et al. observe strong P Cygni emission from the stellar-wind lines C~IV and Si~IV, as well as extremely strong stellar and nebular He~II $\lambda$1640 emission. All efforts to reproduce both the emission and the absorption profiles with standard models failed. The strong emission would require an additional nebular contribution, an unreasonably young stellar population age, and/or a top-heavy IMF, while the absorption requires a much lower than observed metallicity. This, and possibly other star-forming galaxies at high redshift, suggest that our understanding of the UV spectra is still incomplete.

Stellar rotation has been identified as an important missing ingredient in the evolution models (Meynet 2009). Rotation leads to generally higher luminosities and higher effective temperatures for massive stars. This is the result of the larger convective core and the lower surface opacity in the presence of rotation. (Recall that hydrogen is the major opacity source and any decrease of its relative abundance by mixing lowers the opacity and therefore increases the temperature.) This trends is important in all massive stars down to $\sim$20~M$_\odot$, depending on metallicity. The lower the metallicity, the more important is the influence of rotation, which ultimately becomes the dominant evolution driver for metal-free stars. 

Last but not least I emphasize the importance of the currently neglected binary evolution in the models (Eldridge \& Stanway 2009; Vanbeveren 2010). Binary stars experience different evolution from that of single stars if their components interact. Duplicity allows the possibility of enhanced mass loss, mass transfer and other binary specific interactions and hence additional evolutionary pathways. In general, modeling binary systems provides effects similar to including rotation in single star models. Both rotation and binary evolution may change some of the predictions of the next generation of synthesis models.

\end{document}